# Admission and Congestion Control for 5G Network Slicing

Bin Han[1], Antonio De Domenico[2], Ghina Dandachi[2], Anastasios Drosou[3], Dimitrios Tzovaras[3],
Roberto Querio[4], Fabrizio Moggio[4], Ömer Bulakci[5], and Hans D. Schotten[1]

[1]TU Kaiserslautern, Germany; [2]CEA LETI, France; [3]CERTH, Greece; [4]TIM, Italy; [5]Huawei GRC, Germany

*Abstract*—Network Slicing has been widely accepted as essential feature of future 5$^{th}$ Generation (5G) mobile communication networks. Accounting the potentially dense demand of network slices as a cloud service and the limited resource of mobile network operators (MNOs), an efficient inter-slice management and orchestration plays a key role in 5G networks. This calls advanced solutions for slice admission and congestion control. This paper proposes a novel approach of inter-slice control that well copes with existing pre-standardized 5G architectures.

## I. INTRODUCTION

Network slicing, initially proposed by the Next Generation Mobile Networks (NGMN) Alliance [1], is now widely considered as one of the most important key enablers in future 5$^{th}$ Generation (5G) mobile communication networks to support multi-tenancy services. Network slicing allows the mobile network operators (MNOs) to create separated "slices" of network resources, including radio resources, infrastructure resources and virtual resources. These slices can be granted to different network tenants - such as mobile network virtual operators (MNVOs) or service providers - so that every tenant is able to deliver its own services to end customers on a logically independent virtual network.

A typical creation of new network slice begins with the demanding tenant issuing a service request, which is then translated into a slice request for the MNO's consideration for admission. Upon admission, the requested slice is created and granted to the tenant. Once granted, the slice will be continuously maintained and provided with service level agreements (SLAs) guarantees, until it is eventually released after its lifetime established in the SLA or upon the tenant's request. This procedure is accomplished by a specified module for cross-slice management and orchestration (M&O).

This new business case, known as Slice as a Service (SlaaS), is an emerge form of public cloud environment. Distinguished from classic cloud services such as Software as a Service or Infrastructure as a Service, the service delivered in SlaaS, i.e. the slices, can be highly heterogeneous from each other in various features including cost, utility, elasticity, and dynamic behavior. To guarantee the agreed quality of service (QoS) for every granted slice subjected to the overall network resource constraints, the cross-slice M&O needs an advanced policy that autonomously and rationally chooses to accept or to decline every slice requests according to the dynamic resource load. Such a mechanism is supposed to deal with:

- Heterogeneous and flexible slice requirements: for instance, enhanced Mobile Broadband (eMBB) slices will require high bandwith and throughput, while massive Machine-Type Communication (mMTC) slices are demanding ultra-high access capacity but only limited throughput.
- Cross-slice optimization: subjected to a limited resource pool, the MNO needs to optimize its admission strategy in order to jointly maximize the performance criteria such as utility, elasticity, resilience and security across different network slice instances (NSIs) of various types.
- Non-stationary dynamics of service requirements: the statistics of tenant behavior such can dramatically vary over time. For example, demands for short-term network services with high throughput can significantly increase shortly before the New Year compared to the rest periods of a year. The MNO shall be able to adapt its admission strategy timely to such inconsistent service requirements.
- Seamless integration with the (pre-)standardized 5G network architecture(s): practical solutions for slice admission control should be compatible with recent (pre-)standardized network architectures such as those specified by 3GPP and ETSI.

Over the recently past years, a variety of enablers for intelligent network slice admission control have been reported, which apply different approaches including Q- Learning [2], Big Data Analytics [3], neural networks [4], heuristic optimization [5] and game theory [6]. Demonstrations on hardware testbeds have shown that appropriate admission control can make benefits in revenue for the MNO [7] [8]. Initial architectural frameworks for slice admission control in 5G networks have also been proposed [9] [10].

In this paper, we introduce a set of innovative solutions for 5G slice admission and congestion control, which is composed of a 3GPP compatible architectural design for inter-slice M&O enhanced with a set of enablers.

The remainder of this paper is structured as follows: Sec. II proposes our extensions to the current pre-standardized 5G architecture, including a novel M&O layer and a cross-slice orchestration design that enables network function sharing. Sec. III introduces the enabling techniques we apply to support the proposed architectural innovation, including two frameworks and three algorithmic implementations. Then we conclude the paper with Sec. IV.

## II. 5G Inter-slice Management & Orchestration

### A. The 5G Network Management and Orchestration Layer

A number of technical challenges were observed by 5G-MoNArch to solve in order to efficiently support multi-tenancy services in 5G [11]. Following 3GPP guidelines, an M&O system is expected to fill the identified gaps using advanced optimization algorithms and leveraging on virtualization techniques. In addition, End-to-end (E2E) management and orchestration shall be performed in a coordinated manner at different levels, including service, network configuration, virtualization, and transport layers. Therefore, the 5G-MoNArch M&O layer [12] interacts with control and network layers to deploy the required NFs and to configure the appropriate interconnections according to the service and network requirements. Current 5G-MoNArch architecture also takes into account the interactions with the 3GPP Management Entities dedicated to Network management and configuration (3GPP Network Management in Figure 1).

In our E2E Service M&O sublayer, service requirements proposed by tenants are translated into network requirements by the Communication Service Management Function (CSMF). The obtained network requirements are then forwarded to the Network Slice Management Function (NSMF), which is composed by innovative functions addressing the management and orchestration of each slice (cross-domain M&O) and the possible interactions among slices in terms of resources and features sharing (cross-slice M&O) in our proposed architecture.

Provided with network requirements for an incoming service request by the CSMF, the cross-domain M&O identifies the most appropriate candidate from predefined templates for specific well-known or standardized slices, which matches the actual dynamic network requirements the best. By adapting the selected slice template to the requested service, it generates an advanced network slice blueprint with extensions and customizations for the specific deployment. Such a slice blueprint defines the constituents of a NSI in terms of its constituents, e.g. network functions (NFs), connectivity and topology, as well as their configuration.

The issued network slice blueprint is then forwarded to the cross-slice M&O where cross-slice self-organizing management and orchestration (SOMO) algorithms are deployed to make decision of slice admission. Once admitted, a new NSI will be created, activated and continuously maintained. During the maintenance, the cross-domain M&O will execute intra-slice SOMO functions such as resource scaling, NF dynamic deployment, (re-)configuration and troubleshooting, in order to enable slice elasticity and enhance the resilience/security features of the slice. Meanwhile, the cross-slice M&O will take care of the interaction and resource sharing among all deployed NSIs, in order to optimize the overall performances such as elasticity, resilience and security of the entire telco-cloud.

### B. Cross-slice orchestration with shared Network Function

In the context of network slicing, described in the introduction, network resources can be either dedicated to a single Network Slice or shared among different Network Slices. Sharing of network resources can provide a degree of optimization of the network and may be applied depending on the particular use case and the particular network state.

Sharing of network resources represents a key issue in the context of Management and Orchestration, where optimization of resources must be achieved against the constraints imposed by the services to be created. 5G networks are being designed in order to provide the highest degree of flexibility in the service provisioning and delivery. This high level of dynamicity increases even more the optimization challenges of the Management and Orchestration layer.

Figure 2, from [13], represents the two main scenarios related to resource sharing in network slicing:

- More communication services can share the same Network Slice Instance (in the picture, Communication Service Instance 1 and 2 share the Network Slice Instance A)
- More Network Slice Instances can share one or more Network Slice Subnet Instances (in the picture, Network Slice Instance A and B share the Network SubSlice Instance AN-2)

As shown in the picture, Network Slice Instances can span over different network segments (Access, Core and Transport).

According to those two scenarios the Management and Orchestration layer has to support the following use cases:

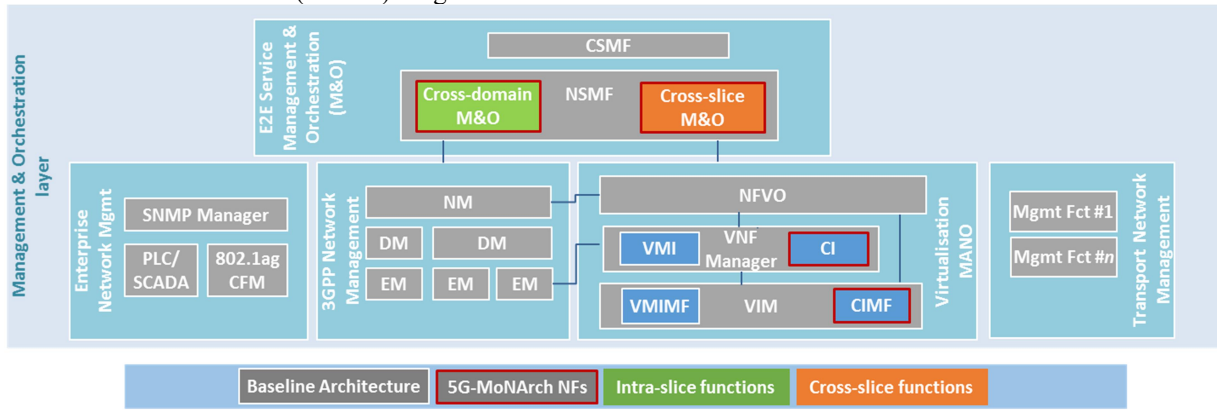

**Figure 1: 5G-MoNArch Management & Orchestration layer**

*1) Network slice allocation using an existing NSI*

The M&O layer has to provide a network slice instance that fits the requested network requirements. The allocation process firstly attempts to re-use an existing NSI in order to optimize the network resources usage. The M&O layer verifies if other existing NSIs, that can support the requested communication service, are available. An identified existing NSI has to be compatible with the network requirements and the sharing must be allowed by the network management policies. The network must continue to support the new overall performance, capacity and lifecycle management requirements for all the communication services it has to provide. If no existing NSI can be used the M&O layer has to create a new one, if possible.

*2) Network slice creation using existing NSSIs*

The M&O layer has to create a new network slice instance (NSI) that meets the requested network requirements. The M&O layer firstly attempts to re-use existing network slice subnets instances (NSSIs), sharing them, in order to optimize the network resources usage. The M&O layer has to identify the constituent network slice subnets that will be used for the network slice. The allocation process verifies, for each requested network slice subnet, if there are sharable NSSIs available that support the requirements, otherwise it has to create a new one, if possible.

*3) Requirements update when the NSI is shared among services*

The M&O layer has to modify a network slice instance (NSI) according to a request for modification of network requirements.

The M&O layer verifies if the current NSI already supports the new requirements and if it still able to supports the new overall performance, capacity and lifecycle management requirements for all the communication services exploiting the NSI. If needed and if it is possible accordingly to the network management policies, the M&O layer reconfigures the NSI, otherwise the operator creates a new network slice instance to support the communication service.

*4) Requirements update when some NSSI is shared among NSIs*

The M&O layer has to modify an existing NSI according to a request for modification of network requirements. Alternatively, if there is another NSI which could support the new network requirements, the M&O layer may decide to use the alternative NSI.

The M&O layer verifies if the current NSI already supports the new requirements. If the NSI doesn't fit the new requirements, The M&O layer evaluates if reconfiguring the current NSI looking for the necessary actions for each constituent NSSI. For each NSSI that has to be modified, the M&O layer evaluates the sharing constraints. After the evaluation process, the M&O layer can update the current NSSI ore create a new one to satisfy the new requirements.

The use cases described in the present chapter imply complex optimization challenges that the Management and Orchestration layer has to address in the context of the lifecycle of Network Slices. In order to achieve this goal, frameworks for Slice admission control and congestion control appear to be pivotal.

### III. SLICE ADMISSION AND CONGESTION CONTROL

The current section discusses on the necessity for a combined approach for controlling the Slice Admission & Congestion. These two concepts are fundamentally connected since they are sequential parts of the same slice management procedure. Once a slice admission has been granted, changes in the behaviour of existing slices may lead to the need for additional network resources.

#### A. Frameworks

*1) Slice Admission Control Framework*

The proposed Slice Admission Control (SLC) Framework consists of a set of Machine Learning technologies that aim to facilitate the resource allocation management through efficient multi-parametric automations in real-time, so as to serve slice admission procedure. Specifically, its role is to analyse the available physical and virtual resources along with their remaining capacity and to decide whether they are capable of accommodating an incoming slice request.

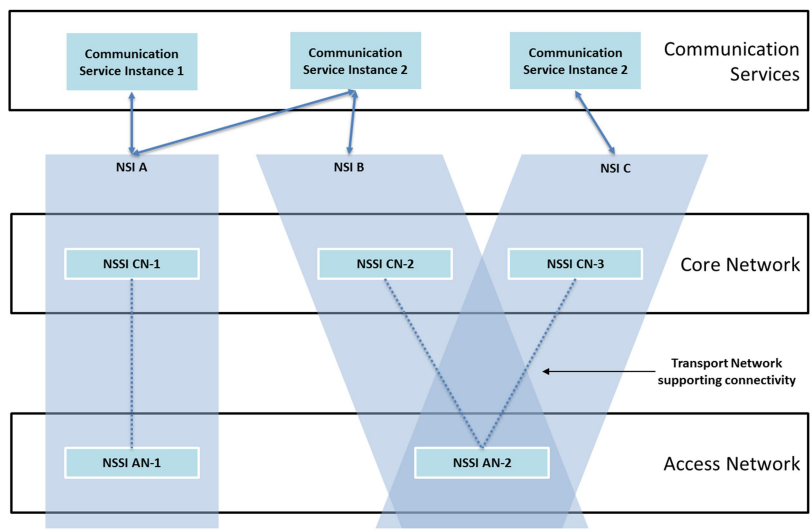

**Figure 2: Slice subnetwork sharing across two or more network slices**

Typical resource allocation management approaches are expected to take into account any slice Service Level Agreements (SLAs), along with all policies they define. Of course, the appropriate Key Performance Indicators (KPIs) are constantly monitored by the MNO. These KPIs can however be *adversarial;* meaning the one may not improve the value of one objective, while no negatively affect the value of another objective, e.g. increasing UE data rates, decreases the cost efficiency of the network operation. Thus, in order to efficiently tackle conflicting problems of a multi-objective nature, multi-objective optimization methods [14] can be utilized, to produce a set of *Pareto*-optimal solutions that abide to specific constraints. This way, MNOs are allowed to interactively select between a number of trade-offs between the multiple objectives [15]. Last but not least, after the admission of a new network slice instance (NSI), the implementation of such SLCs must additionally ensure that the resource allocation methods can optimise the network utilization, while also meeting the SLAs of each NSI.

Within our architecture, slice resource consumption is monitored and managed in the end-to-end (E2E) Service Management & Orchestration (M&O) sublayer and more specifically by a module for resource orchestration in the Cross Slice Self-Organising M&O functions. In this context, the architectural diagram of the proposed approach is shown in Figure 3. Given the aforementioned factors as input, as well as the resource orchestration module, the framework decides if the new NSI can be deployed or not.

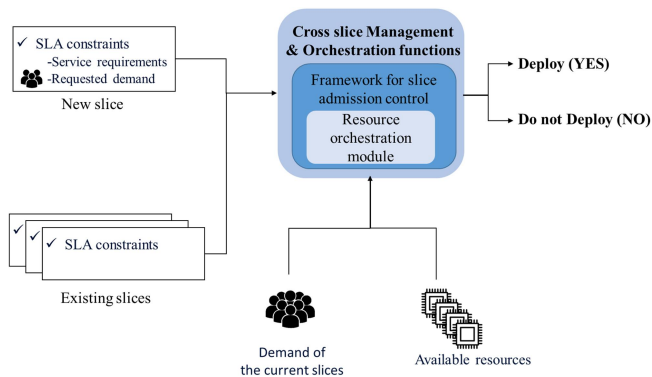

**Figure 3: Architectural diagram of the proposed framework for slice admission control**

*2) Slice Congestion Control Framework*

After a slice admission, the need for additional network resources is related to the arrival of a new slice demand or to perceived performance reduction at the slice level, due, for instance, to an increased slice load. However, these slices have different requirements and priorities where elastic slices do not require fully dedicated resources and have relaxed security and reliability constraints. On the other hand, higher priority slices have more stringent requirements in terms of allocated resources and reliability.

This difference of slices requirements can be useful in the case of high load where congestion control strategy are applied on elastic slices so that higher priority slices are accepted. Therefore, the congestion control framework has to 1) identify the slices with looser requirements for which the amount of allocated resources can be reduced and 2) update their resource allocation.

The Cross-slice Congestion Control (CSCC) function shown in Figure 4 may decide, based on resources availability, slices requirements, and the queue state, to scale down the allocated resources to one or multiples slices in order to accept a larger number of slices of higher priority. The proposed CSCC has to be able to foresee the impact of such decision on the overall system performance [16]. This intelligence is ensured by using reinforcement learning (RL) techniques that allow to make the optimal decisions jointly maximizing resources utilization and guaranteeing that available resources are allotted according to the slice priorities [17].

The CSCC will be implemented as an additional function on the top of the admission control framework at the orchestrator level. In particular, it is associated to the cross Slice SOMO module inside the NSMF [12] as it enables to manage virtual resources across different slices such that the overall resource utilization efficiency is maximized, dropped slices are minimized, and the requirements of accepted slices are satisfied.

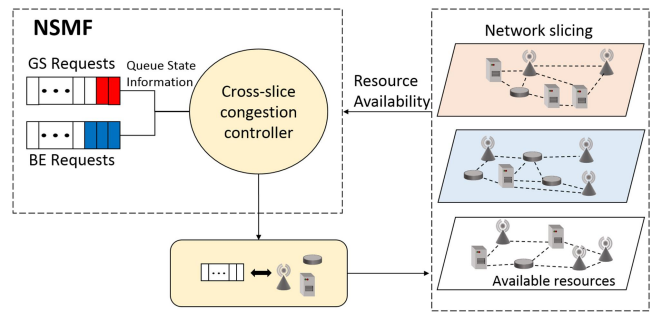

**Figure 4: Cross-slice admission and congestion control framework**

*B. Implementations and Demonstrations*

*1) Slice Admission Control*

Some indicative outcomes of the Slice Admission Control Framework (see Section III.A.1) are illustrated in Figure 5. Without loss of generality, the computational resources considered are labelled as cpuX (X:{1:4}). As it can be seen in the given scenario, each slice is executed on a subset of the computational resources (in this case one CPU) for a specific period (i.e. from 09:00 to 17:00 on the 4th of Dec), as also indicated by the results of the multi-objective resource orchestration manager. The execution of a certain slice can theoretically be split among several cpus; however, it is has been set as a hard constraint in the demonstrated example.

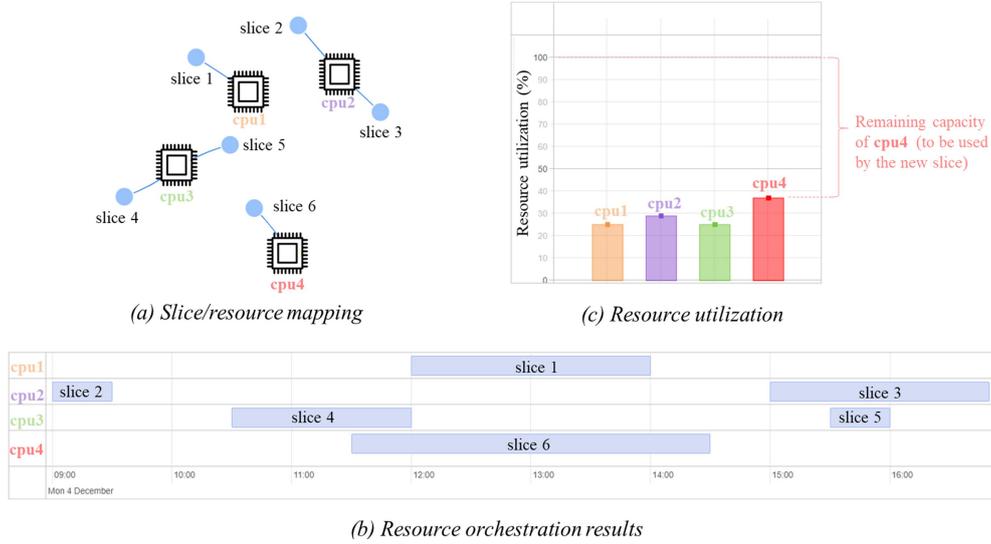

*(a) Slice/resource mapping*

*(c) Resource utilization*

*(b) Resource orchestration results*

**Figure 5:** Results of analysis of the capacity of the computational resources and their availability to accommodate a new slice

Specifically, an indicative mapping of the slices to the computational resources is demonstrated in Figure 5a, i.e., slices 2 and 3 are executed on cpu2, and thus, connected by an edge. Figure 5b presents the actual mapping of the slices to the computational resources over time as computed by the resource orchestration algorithm, i.e., slices 2 & 3 are executed on cpu2 from 9:00-9:30 and from 15:30-16:00, respectively.

Finally, Figure 5c illustrates the resource utilization for each computational resource, and the remaining capacity that can be utilized by the new slice. All the resources are less than 50% occupied, and given the resource demand of the new NSI (e.g. 20% CPU power) there is enough remaining capacity to accommodate its efficient deployment.

*2) Q-Learning Assisted Cross-slice Congestion Control*

At this stage, both admission and congestion control decisions are taken by the framework, and two slice classes are defined: best effort (BE) and guaranteed service (GS). In order to prioritize the deployment of GS requests, a higher reward is assigned for accepting their requests. It is important to note also that negative rewards will be considered when dropping a GS request so that the policy is pushed toward deploying more GS requests rather than BE slices. In this first study Q-learning is used to learn to optimal strategy to implement at the CSCC [17]. In the future months, more complex algorithms will be implemented that can take into account more realistic environment. In the results shown in Figure 6, the proposed solution is compared with a greedy policy in term of accepted and dropped slice requests. The results show that the proposed solution is able to improve the resource utilization enabling to increase the percentage of accepted slice request without negatively affecting the performance at the GS slices.

*3) Genetic Slice Admission Strategy Optimizer*

While the Q-Learning methods exhibits good effectiveness in optimizing admission strategies, its performance can highly depend on the initialization. An inappropriate specification of the initial Q-matrix can lead to slow or premature convergence.

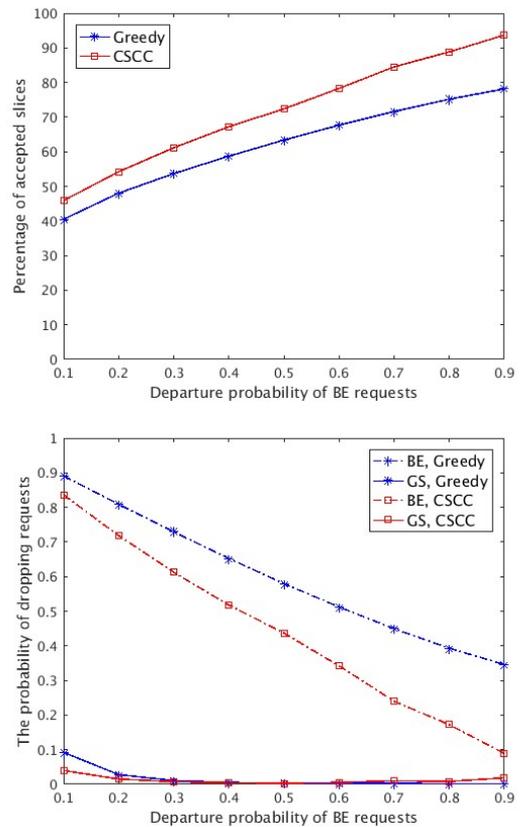

**Figure 6:** Percentage of accepted slices for Greedy and the proposed QL-based CSCC as a function of BE departure probability (top), and dropping probability for Greedy and QL-based CSCC as a function of BE departure probability (bottom)

Furthermore, it may fail to adapt sufficiently fast when the environment is not consistent but highly dynamic. A straightforward solution to overcome these drawbacks is to train and periodically retrain the Q-Learning system with carefully selected sets of reference strategies.

As a good complement to Q-Learning, genetic algorithm (GA) provides an efficient evolutionary method to generate sub-optimal strategy sets with simple and parallelizable implementation [18]. In GA, each individual strategy is encoded into a unique binary sequence. By recursively applying the random operations of reproduction, crossover and mutation on the candidate strategy set, which is known as the "population", GA drives the entire population to evolve in a windingly way towards the optimum. Generally, GA exhibits the merits of modeless implementation, fast convergence, robustness to inconsistency, and high scalability. Some sample outputs of GA-based strategy optimization are illustrated in Figure 7 [18].

## IV. CONCLUSION

In this paper, we have introduced an innovative set of enablers and algorithmic designs for intelligent slice admission and congestion control in future 5G networks. The proposed architecture is capable for coherence with existing ETSI and 3GPP architectures, making it competitive in 5G pre-standardization. The algorithms are demonstrated as effective by numerical simulations.

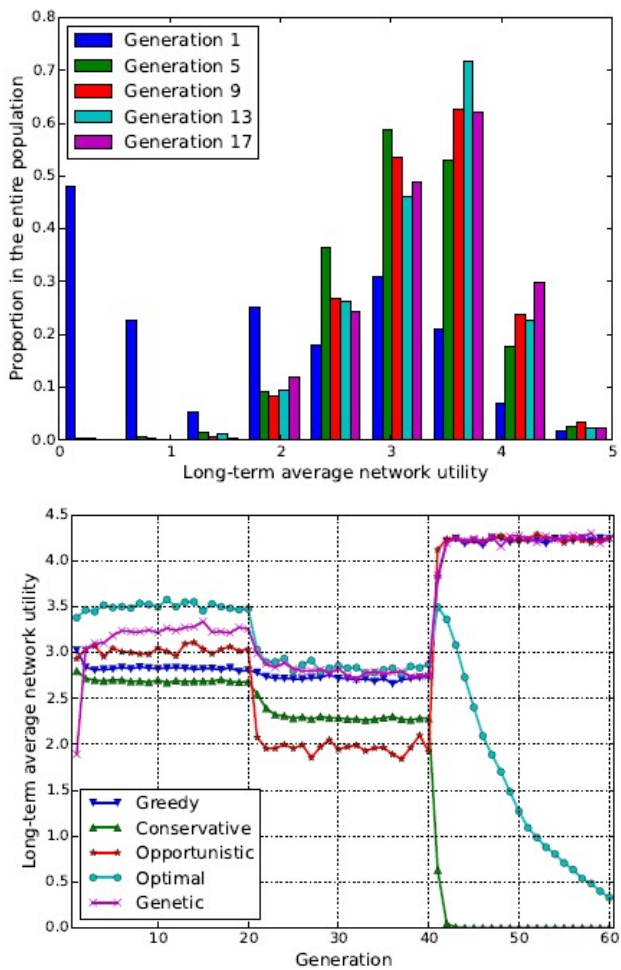

**Figure 7: The evolution of GA population towards an optimal long-term normalized network utility rate (top), and GA's robustness to inconsistent environment in comparison to static benchmark strategies (bottom)**

ACKNOWLEDGMENT

This work has been performed in the framework of the H2020 project 5G-MoNArch co-funded by the EU. The views expressed are those of the authors and do not necessarily represent the project. The consortium is not liable for any use that may be made of any of the information contained therein.